\documentclass[pdflatex,sn-nature]{sn-jnl}

\usepackage{graphicx}
\usepackage{amsmath,amssymb}
\usepackage{bm}
\usepackage{xcolor}
\usepackage{siunitx}
\usepackage{booktabs}
\usepackage{placeins}
\usepackage{soul}
\usepackage{mathalpha}
\usepackage{changes}

\begin{document}

\title[Fractality-induced photonic topological insulators]{Fractality-induced photonic topological insulators}

\author[1]{\fnm{Shuming} \sur{Zhang}}\equalcont{These authors contributed equally to this work.}
\author[1]{\fnm{Zhaoxin} \sur{Wu}}\equalcont{These authors contributed equally to this work.}
\author[2]{\fnm{Lumen} \sur{Eek}}
\author*[2]{\fnm{Cristiane} \sur{Morais Smith}}\email{c.demoraisSmith@uu.nl}
\author*[1]{\fnm{Zhaoju} \sur{Yang}}\email{zhaojuyang@zju.edu.cn}

\affil[1]{\orgdiv{School of Physics and Zhejiang Key Laboratory of Micro-nano Quantum Chips and Quantum Control},
\orgname{Zhejiang University},
\orgaddress{\city{Hangzhou}, \postcode{310058}, \state{Zhejiang Province}, \country{China}}}

\affil[2]{\orgdiv{Institute of Theoretical Physics},
\orgname{Utrecht University},
\orgaddress{\city{Utrecht}, \postcode{3584 CC}, \country{Netherlands}}}

\abstract{Fractal lattices have recently emerged as a promising setting for topological wave physics, but in most realizations the topological character is inherited from externally engineered couplings, gauge fields, or temporal modulation rather than from the fractal geometry itself. Here, we experimentally realize a photonic higher-order topological insulator in which the topology is induced solely by the self-similar geometry of a Sierpiński-gasket lattice. Following the isospectral reduction method recently proposed by Eek \textit{et al.}~\cite{Eek2025}, we show that the fractal waveguide array with uniform nearest-neighbor couplings can be mapped onto an effective breathing Kagome model that supports corner states. We selectively excite these modes with a weakly coupled detuned auxiliary waveguide and directly observe robust corner localization in real space, whereas an otherwise equivalent uniform triangular lattice exhibits only bulk diffraction under the same protocol. Spectral analysis and open-boundary calculations associate the observed states with nontrivial $C_3$ rotational topology, and disorder measurements further show that the corner localization persists over a finite range of random and symmetry-preserving disorder. Our results establish fractal geometry itself as a mechanism for generating topological boundary states in photonic lattices.
}

\maketitle

\section*{Introduction}
Topological insulators~\cite{Hasan2010,Qi2011} are phases of matter distinguished not by spontaneous symmetry breaking or local order parameters, but by global topological invariants encoded in their wave functions, which give rise to robust boundary states through the bulk-boundary correspondence \cite{hatsugai1993chern}. Since their original formulation in electronic condensed-matter systems, topological phases have profoundly reshaped our understanding of band theory and quantum matter, revealing that insulating bulks can support protected conducting states immune to disorder, imperfections, and local perturbations \cite{kane2005quantum,bernevig2006quantum,konig2007quantum}. This paradigm has subsequently extended far beyond solid-state materials and has found an especially versatile arena in photonics, where synthetic lattices, reconfigurable geometries, engineered couplings, and direct imaging of wave evolution provide exceptional control over topological phenomena. Over the past decade, topological photonics~\cite{Ozawa2019} has enabled the observation of the unidirectional edge transport~\cite{Haldane2008, Wang2009, Fang2012, Rechtsman2013, Hafezi2013, Khanikaev2013, Lu2014, Wu2015} and more recently higher-order topological corner modes~\cite{benalcazar2017quantized,serra2018observation,peterson2018quantized,imhof2018topolectrical,chen2019direct,mittal2019photonic,el2019corner,li2020higher,xie2021higher}, thereby establishing light-based platforms as a powerful testbed for exploring topological wave physics in regimes difficult to access in electronic systems.

In parallel, increasing attention has turned to geometric complexity~\cite{kempkes2019design,vardeny2013optics,kollar2019hyperbolic,yu2020topological,zhang2022observation} as a new degree of freedom for topological design. Among the most intriguing settings are fractal lattices~\cite{Mandelbrot1967}, whose self-similar architecture and noninteger Hausdorff dimension fundamentally challenge the conventional crystalline framework underlying most topological classifications. Fractal systems have long been associated with unusual spectral and transport properties, including localization, anomalous diffusion, and critical wave dynamics \cite{orbach1986dynamics,van2016quantum,mirlin2006exact,yao2019critical,xu2021quantum}, and they have recently emerged as a fertile platform for topological physics \cite{manna2022higher, manna2023inner, osseweijer2024haldane, Salib2024} across electronic \cite{canyellas2024topological} and classical-wave~\cite{yang2020photonic, Tobias2022,zheng2022observation, Junkai2022,li2023fractality,li2023light,zhong2024light} systems. In particular, photonic fractal structures have been shown to host unconventional topological effects such as Floquet topological transport~\cite{Tobias2022,li2023light} and higher-order boundary localization~\cite{zhong2024light} in self-similar lattices. Yet, in nearly all such realizations, the topological character is inherited from artificial gauge fields, engineered couplings, or parent Hamiltonians that are already topologically nontrivial before being embedded into a fractal geometry. This raises a more fundamental possibility: whether fractal geometry can itself serve not merely as a substrate that reshapes topology, but as an intrinsic mechanism that generates it. Very recently, theory has suggested that self-similar fractal lattices may indeed induce topological corner states even in otherwise topologically trivial models, pointing to a qualitatively new route toward topology in non-crystalline systems \cite{Eek2025}. Together, these developments place fractal topological photonics at the frontier of geometry-driven wave engineering and motivate the search for genuine topological phases induced purely by fractality.

Here, we experimentally realize fractality-induced higher-order topology in a genuine photonic Sierpiński-gasket lattice fabricated by femtosecond-laser direct writing. Guided by the recent isospectral reduction picture, we implement a self-similar waveguide array with uniform nearest-neighbor couplings and nearly uniform onsite potentials, such that the observed topological response originates from the fractal geometry itself rather than from artificial gauge fields or pre-engineered dimerization. To directly probe the predicted corner modes, we introduce an auxiliary waveguide with a tunable detuning that enables state-selective excitation near a target corner-state energy. We show, through both numerical simulations and optical measurements, that light injected under this protocol remains strongly confined to the corners of the fractal lattice, whereas in a topologically trivial uniform triangular lattice it rapidly diffracts into the bulk. Furthermore, by combining spectral analysis, real-space imaging, and disorder tests, we demonstrate that these corner-localized states are associated with nontrivial rotational topology and remain robust against considerable fully random and $C_3$-symmetric disorder. Our results provide, to our knowledge, the first photonic observation of topology generated purely by fractality, and establish fractal photonic lattices as a platform for geometry-driven topological control beyond crystalline settings.

\begin{figure}[t]
    \centering
    \includegraphics[width=0.85\linewidth]{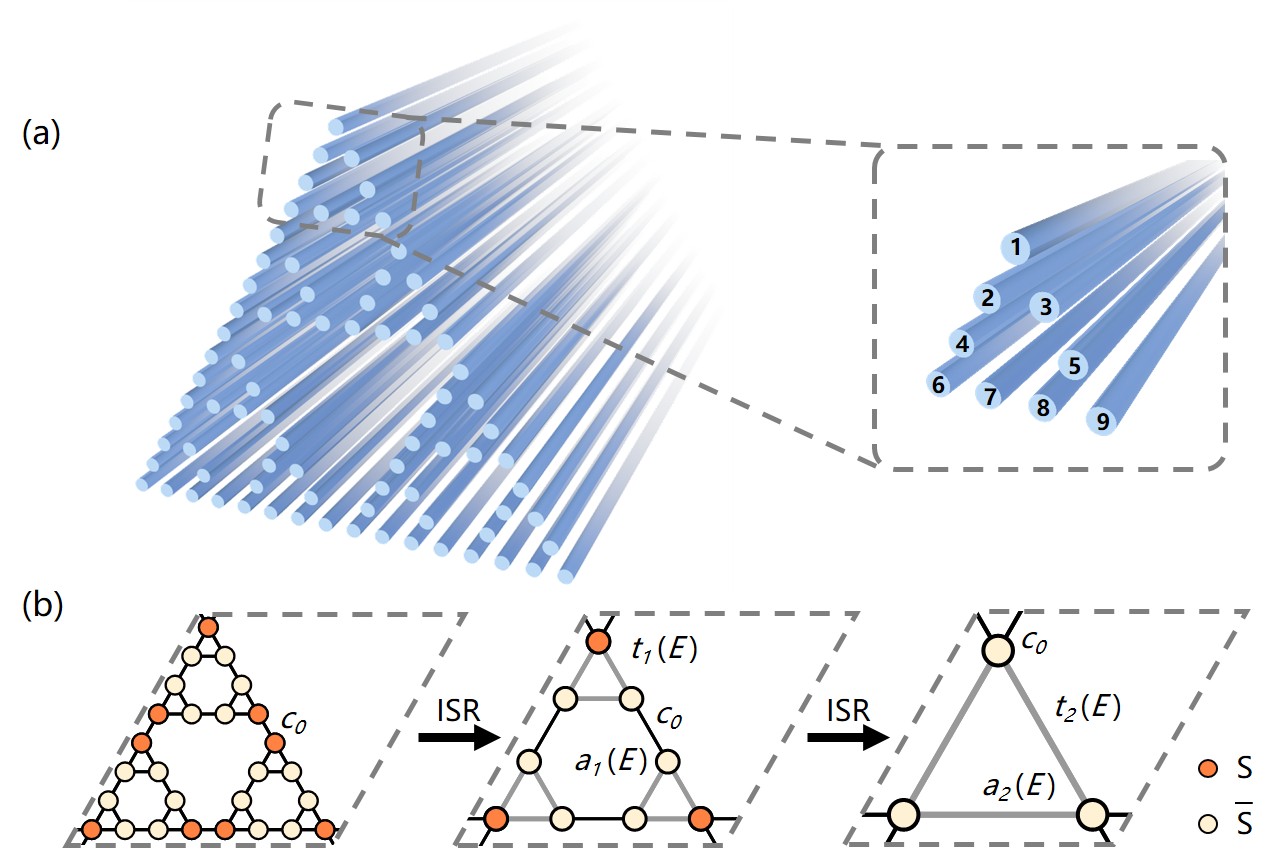}
    \caption{Sierpiński-gasket lattice consisting of optical waveguides. (a) Schematic of the third-generation fractal photonic lattice. The inset shows the first-generation Sierpiński-gasket lattice. (b) Isospectral reduction of the Sierpiński-gasket lattice. Orange circles indicate the sublattice sites $S$ that were retained at each step, while the yellow circles indicate the eliminated subset $\bar{S}$. Gray (black) bonds denote effective intracell (intercell) couplings.}
    \label{fig1}
\end{figure}

\section*{Results}
\textbf{Model}. We study a third-generation Sierpiński-gasket waveguide lattice [Fig.~\ref{fig1}(a)] fabricated by femtosecond-laser direct writing in glass~\cite{Rechtsman2013} (see Sec. I of the Supplementary Information). The underlying fractal has a Hausdorff dimension \(d_H=\log_2 3 \approx 1.585\), and is generated by recursively removing the central triangular site, thereby producing a self-similar lattice hierarchy. Under the paraxial approximation, light propagation along the waveguides is governed by
\begin{equation}
i\frac{\partial \psi_m}{\partial z}
= \sum_{\langle m,n\rangle} c_{m,n}\psi_n + \varepsilon_m \psi_m,
\end{equation}
where \(\psi_m\) is the field amplitude at the \(m\)th waveguide, \(c_{m,n}\) denotes the nearest-neighbor coupling, and \(\varepsilon_m\) is the onsite detuning. Here \(z\) plays the role of an effective time coordinate. To isolate the role of geometry, we engineer a lattice with uniform nearest-neighbor couplings \(c_{m,n}=c_0=2\,\mathrm{cm}^{-1}\) and vanishing onsite detuning, \(\varepsilon_m=0~\text{cm}^{-1}\), throughout the fractal optical array~\cite{Tobias2022,li2023light}.

The topological origin of this lattice follows from the isospectral-reduction framework recently developed for fractal systems~\cite{Eek2025}. As shown in Fig.~\ref{fig1}(b), the lattice is partitioned into a retained subset $S$ (orange) and an eliminated subset $\bar{S}$ (yellow), yielding an energy-dependent effective Hamiltonian defined on $S$, with energy-dependent hopping $t_n(\beta)$ and onsite potential $a_n(\beta)$. Repeated application of isospectral reduction maps the higher-generation Sierpiński-gasket lattice onto an effective breathing Kagome lattice, with the self-similar geometry encoded in renormalized hopping and onsite terms. Within this reduced description, corner states arise when the eigenenergy satisfies $\beta_c=a_n(\beta_c)$, and the effective breathing ratio lies in the topological regime, $t_n(\beta_c)<c_0$ (equivalently, the intracell coupling is smaller than the intercell one, see Sec. IV.B of the Supplementary Information for more details). Fractality thus does not merely reshape a preexisting topological phase, but generates higher-order topology from an otherwise topologically trivial lattice.

\begin{figure*}[ht]
    \centering
    \includegraphics[width=1\linewidth]{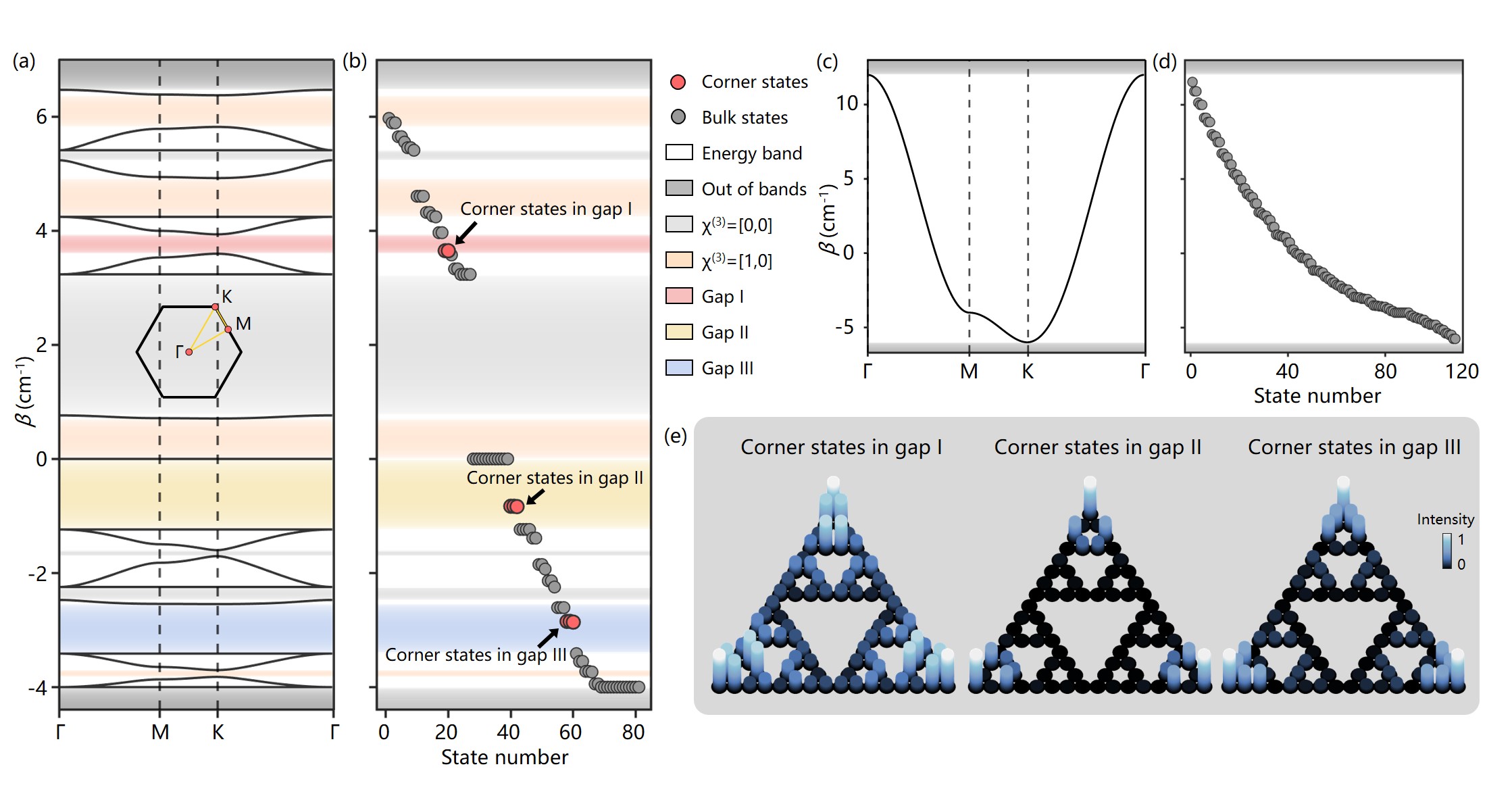}
    \caption{Fractality-induced higher-order topology.
(a) Band structure of the second-generation Sierpiński-gasket lattice under the periodic boundary condition (PBC), which is used to characterize the bulk topology associated with the third-generation fractal lattice under the open boundary condition (OBC).
Topologically nontrivial band gaps with nonzero rotational invariants $\chi^{(3)}$ are highlighted in red, yellow, blue, and orange, whereas topologically trivial gaps are shaded in gray. 
Inset: the first Brillouin zone of the Sierpiński-gasket lattice and the underlying triangular lattice.
(b) Energy spectrum of the third-generation Sierpiński-gasket lattice under OBC. 
Bulk states and topological corner states are denoted by gray and red circles, respectively. 
(c) Energy band structure of a uniform triangular lattice with a single-site unit cell, shown for comparison, exhibiting a topologically trivial spectrum.
(d) Energy spectrum of the uniform triangular lattice under OBC.
(e) Intensity distributions obtained by summing the normalized eigenstate intensities of the corner states within the three topological gaps I--III.
}
    \label{fig2}
\end{figure*}

We next calculate the band structure of the photonic Sierpiński-gasket lattice and evaluate the corresponding $C_3$ symmetry indicator $\chi^{(3)}=[K_1^{(3)}, K_2^{(3)}]$ for the relevant gaps (see Sec. III of the Supplementary Information).  Under periodic boundary conditions, the fractal lattice exhibits multiple gapped band manifolds [Fig.~\ref{fig2}(a)], among which the topological and trivial gaps are marked in bright colors and gray, respectively. By contrast, the uniform triangular lattice [Fig.~\ref{fig2}(c)] supports only a single topologically trivial band. To reveal their properties in a finite size, we further investigate the fractal and uniform lattices under open boundary conditions. For the fractal lattice [Fig.~\ref{fig2}(b)], several in-gap eigenstates, highlighted by red circles, appear inside the gaps with nontrivial rotation topology, $\chi^{(3)}=[1,0]$. Their eigenvalues satisfy the condition $\beta_c=a_n(\beta_c)$ and $t_n(\beta_c)<c_0$, with eigenstates concentrated at the three corners [Fig.~\ref{fig2}(e)], identifying them as topological higher-order corner states. In contrast, the uniform triangular lattice [Fig.~\ref{fig2}(d)] hosts no analogous in-gap corner modes, and all eigenstates remain extended over the bulk. This contrast demonstrates that the higher-order corner states originate from the fractal geometry itself.

In view of the above discussion, we stress that a nontrivial gap under periodic boundary conditions does not by itself guarantee corner states in a finite sample. Their emergence additionally requires the isospectral-reduction confinement condition, namely that the corner-state energy satisfies the condition that the corresponding effective hopping lies in the topological regime $t_n(\beta_c)<c_0$ (see Sec. IV of the Supplementary Information). 

\begin{figure*}[htbp]
    \centering
    \includegraphics[width=1\linewidth]{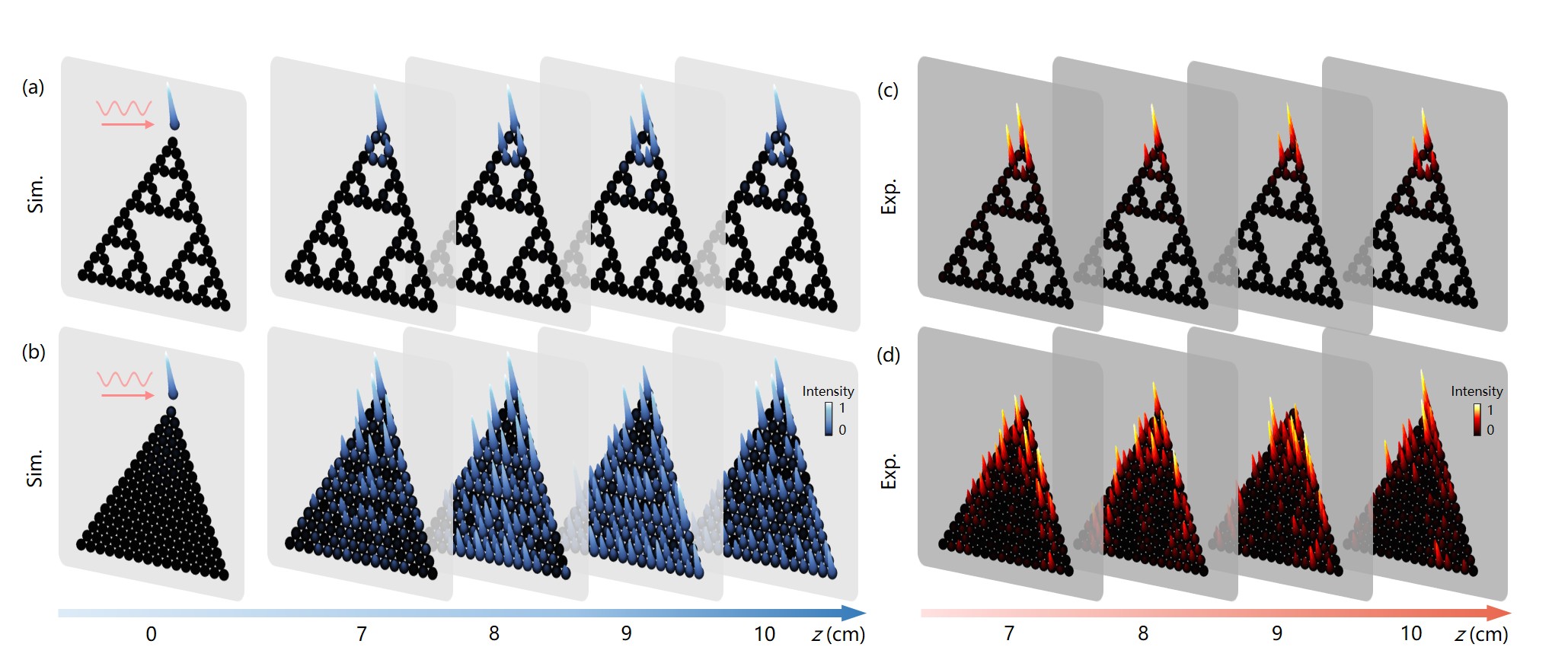}
    \caption{Excitation of topological corner states with an auxiliary waveguide.
(a, b) Numerically simulated evolution of the optical intensity distribution in a fractal Sierpiński-gasket lattice (a) and a uniform triangular lattice (b), both excited via an identical auxiliary waveguide with $\varepsilon_{aux}\approx-0.84\, \text{cm}^{-1}$.
In the fractal lattice, the optical field remains strongly localized at the corner throughout propagation, whereas in the uniform triangular lattice, the excitation rapidly spreads into the bulk.
(c, d) Experimentally measured output intensity distributions recorded at different lattice facets for the fractal lattice (c) and the uniform triangular lattice (d), respectively.
}
    \label{fig3}
\end{figure*}

\textbf{Experimental realization}. Having established the existence and topological origin of the corner modes from the bulk spectra and open-boundary eigenstates, we now turn to their state-selective excitation and direct real-space observation in the photonic lattice. To selectively excite the predicted corner modes, we employ an auxiliary waveguide~\cite{noh2018Topological} weakly coupled to the top-corner site of the fractal lattice. In experiments, the 635-nm light is injected into this auxiliary channel, while its onsite detuning $\varepsilon_{\rm aux}$ is tuned such that the corresponding propagation constant is resonant with the target corner-state eigenenergy. The auxiliary coupling is chosen to be sufficiently weak (\(c=0.6~\mathrm{cm}^{-1}\)) so that the intrinsic spectral properties of the fractal lattice remain nearly unperturbed. In this way, the auxiliary waveguide acts as an external drive that prepares a corner-state-selective initial excitation without introducing artificial hopping modulation into the lattice itself. Once most of the optical power has been transferred into the lattice, the auxiliary channel is effectively truncated, and the subsequent propagation is governed solely by the Hamiltonian of the fractal array. Because the auxiliary waveguide couples only to the selected corner, this protocol predominantly addresses the corner-state manifold localized near that corner. More details can be found in Sec. II.A of the Supplementary Information.

We then track the propagation dynamics by performing sliced measurements along the longitudinal direction \(z\). Figs.~\ref{fig3}(a) and \ref{fig3}(b) show the corresponding numerical evolutions for the fractal Sierpi\'nski-gasket lattice and, for comparison, a uniform triangular lattice with otherwise identical coupling parameters. The simulated intensity profiles are obtained from the paraxial propagation $\psi(z)=U(z)\psi(0)$ with $U(z)=\exp(-iHz)$, where $H$ is the effective tight-binding Hamiltonian of the waveguide array and $\psi(0)$ is the prepared initial state. To prepare the initial states, 635-nm light is injected into the auxiliary waveguide at \(z=0~\text{cm}\), whose onsite detuning is set to \(\varepsilon_{\rm aux}\approx -0.84~\mathrm{cm}^{-1}\), resonant with the corner-state energy in gap II. After the auxiliary section is terminated at \(z=4.4~\mathrm{cm}\), the subsequent evolution is governed solely by the lattice Hamiltonian, allowing the prepared excitation to probe the intrinsic dynamics of the target corner-state manifold.

Under these conditions, the fractal lattice exhibits a strikingly different evolution from that of the topologically trivial control lattice. In the Sierpi\'nski-gasket lattice, the optical field remains strongly concentrated near the selected corner throughout propagation, with the dominant intensity residing on the sites carrying the largest corner-state weight. In contrast, for the uniform triangular lattice under the same excitation protocol, the injected light rapidly spreads away from the corner and diffracts into the bulk, with no comparable corner confinement. The experimentally measured intensity distributions at different propagation lengths [Figs.~\ref{fig3}(c) and \ref{fig3}(d)] closely reproduce the simulated evolution in both systems, establishing a clear distinction between the fractal and trivial lattices. Note that the same excitation strategy can also access the localized corner states in gap III (see Sec. II.A of the Supplementary Information), whereas the weakly localized mode in gap I is not cleanly resolved experimentally because the gap size is relatively small and the prepared initial state has non-negligible overlap with bulk modes. These observations provide direct evidence that the observed localization is not a generic boundary effect of the input geometry, but arises from the fractality-induced topological corner states predicted by the spectral analysis.

\begin{figure*}[htbp]
\centering
\includegraphics[width=1\linewidth]{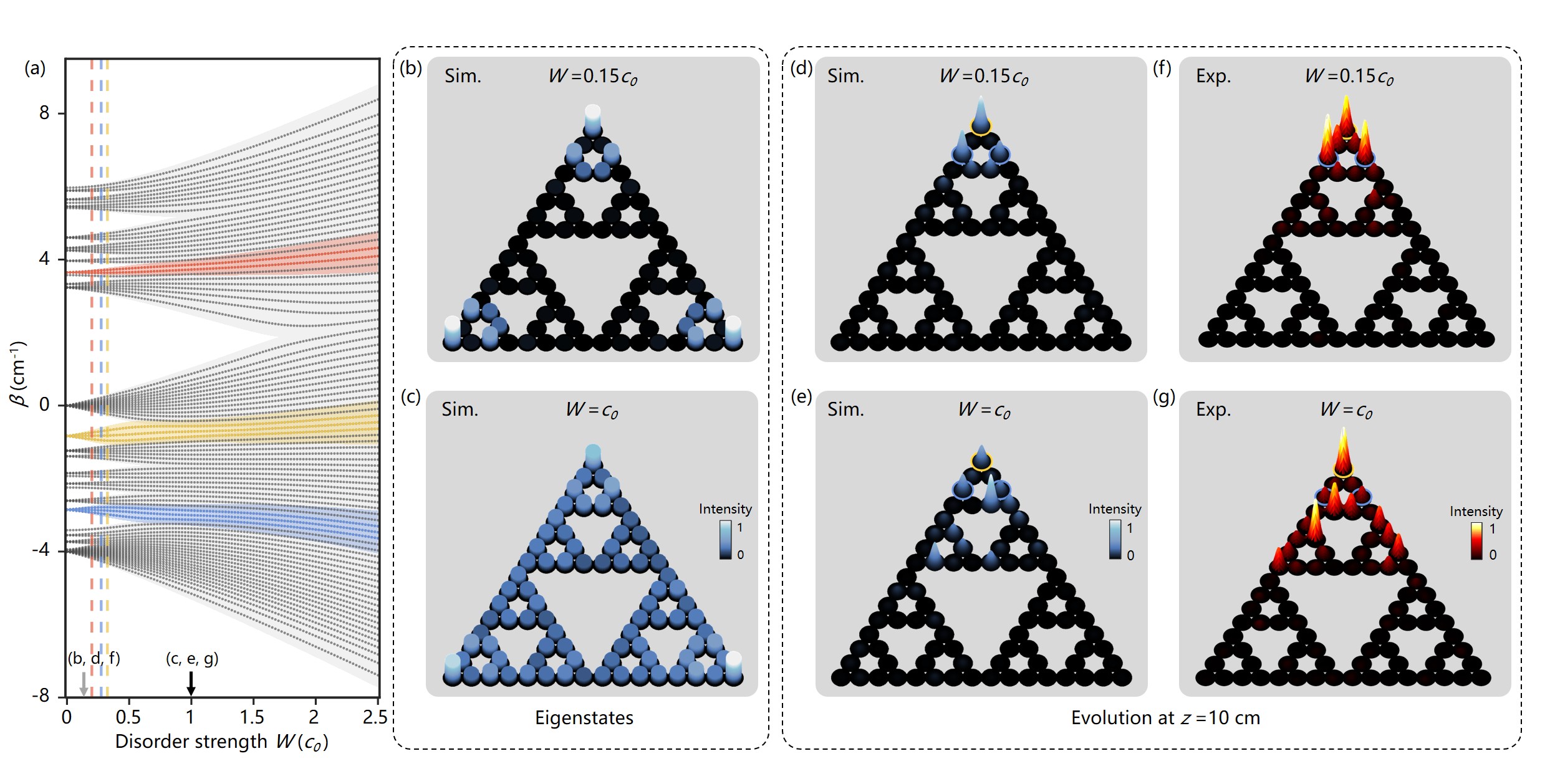}
\caption{Topological robustness against random onsite disorder.
(a) Disorder-averaged spectrum under OBC as a function of the disorder strength $W$. The corner states in gaps I--III are highlighted in red, yellow, and blue, respectively, while the remaining states are shown in gray. The colored vertical dashed lines mark the corresponding disorder strengths at which the gaps close. The gray/black arrows indicate the two disorder strengths used in (b)--(g).
(b, c) Simulated disorder-averaged eigenstate intensity distributions of the corner states in gap II for $W=0.15c_0$ and $W=c_0$.
(d)--(g) Simulated and experimental output intensity distributions at $z=10~\mathrm{cm}$ under three-site excitation for single disorder realizations, with $W=0.15c_0$ in (d, f) and $W=c_0$ in (e, g).}
\label{fig4}
\end{figure*}

\textbf{Topological robustness}. With the selective excitation and real-space localization of the corner modes established, we finally investigate their robustness against onsite disorder. We first consider the random onsite disorder uniformly distributed in $[-W,W]$. This disorder is experimentally realized through varying the laser writing speed~\cite{Sun2024PhotonicFloquet,Hou2026QuantumLight}. Under this condition, the rotational invariant $\chi^{(3)}$ is no longer well defined. Nevertheless, the persistence of the corner states can still be directly examined from the open-boundary spectrum, eigenstate profiles, and propagation dynamics.

Fig.~\ref{fig4}(a) shows the disorder-averaged open-boundary spectrum as a function of the normalized disorder strength $W$, obtained over $2\times10^3$ disorder realizations. The corner states within gaps I--III are highlighted in red, yellow, and blue, respectively, while the remaining states are shown in gray. The corresponding shaded regions indicate the standard deviation over disorder realizations. In the weak-disorder regime, the corner states remain spectrally isolated within the gaps. As the disorder strength increases, the gaps gradually close at \(W_1=0.2c_0\), \(W_2=0.32c_0\), and \(W_3=0.27c_0\) for gaps I, II, and III, respectively. Beyond these critical values, the corner states merge into the bulk continuum, signaling the breakdown of topological protection. To visualize this transition directly, we compare the disorder-averaged intensity profiles of the corner states in gap II in the weak- and strong-disorder regimes [Figs.~\ref{fig4}(b) and \ref{fig4}(c)], respectively. At \(W=0.15c_0\), the eigenstate intensity remains strongly localized at the three corners, whereas at \(W=c_0\) it becomes spatially extended across the lattice. In the experiment, we adopt a three-site excitation scheme (see Sec.~II.B of the Supplementary Information) and record the output field at \(z=10~\mathrm{cm}\). For weak disorder, the simulated and measured output preserves pronounced corner localization [Fig.~\ref{fig4}(d) and ~\ref{fig4}(f)]. By contrast, for strong disorder, both output fields spread into the bulk [Figs.~\ref{fig4}(e) and \ref{fig4}(g)]. These results demonstrate that the corner states remain robust against weak random disorder, but lose their topological protection once the relevant gap closes.

We further analyze a $C_3$-symmetric disorder configuration, for which the rotational invariant $\chi^{(3)}$ remains well defined (see Sec.~V of the Supplementary Information). We note that the $C_3$-symmetric disorder exhibits a higher robustness than the fully random disorder considered here, reflecting the additional protection from the preserved rotational symmetry.
Together, these results show that the fractality-induced topological corner states remain robust against considerable onsite disorder, while sufficiently strong disorder eventually destroys their spectral isolation and real-space localization.~\\

\textbf{Conclusion}. In summary, we have experimentally realized fractality-induced higher-order topology in a genuine photonic Sierpiński-gasket waveguide lattice. Guided by the isospectral-reduction framework, we have confirmed the recent theoretical prediction \cite{Eek2025} that the self-similar fractal geometry maps onto an effective breathing Kagome model and can generate topological corner states without artificial gauge fields, staggered couplings, or temporal modulation. By combining bulk spectral analysis, open-boundary calculations, state-selective excitation through an auxiliary waveguide, and real-space measurements, we directly observed corner-localized modes that are absent in a topologically trivial triangular lattice. We further demonstrated that these modes remain robust over a finite range of fully random onsite and \(C_3\)-symmetric disorder, confirming the stability of the fractality-induced topological phase.

Beyond establishing a photonic platform for topology in non-crystalline geometries, our work points to a broader route toward geometry-driven wave control in self-similar systems. Because the present femtosecond-written waveguide architecture is naturally compatible with nonlinear and quantum-optical implementations, it may provide a versatile starting point for exploring the interplay of fractal topology with Kerr-type nonlinear dynamics~\cite{maczewsky2020nonlinearity,jurgensen2021quantized,jurgensen2023quantized,hou2026nonlinearphotonictripartitephase} and quantum interference~\cite{ehrhardt2024topological,selim2025selective,gao2024Quantum}. These possibilities establish self-similar photonic lattices as a promising setting for connecting fractal topology with nonlinear wave dynamics and integrated quantum optics.~\\

\backmatter

\bmhead{Supplementary information}
The Supplementary Information is available for this paper.

\bmhead{Acknowledgements}
This research is supported by the National Key R\&D Program of China (Grant No. 2023YFA1406703, 2022YFA1404203) and the Fundamental Research Funds for the Central Universities (Grant No. 226-2025-00124). L.E. and C.M.S. acknowledge the research program “Materials for the Quantum Age” (QuMat) for financial support. This program (registration number 024.005.006) is part of the Gravitation program financed by the Dutch Ministry of Education, Culture and Science (OCW).

\bmhead{Competing interests}
The authors declare no competing interests.






\end{document}